
\def\tenpoint{\normalbaselineskip=12pt plus 0.1pt minus 0.1pt
  \abovedisplayskip 12pt plus 3pt minus 9pt
  \belowdisplayskip 12pt plus 3pt minus 9pt
  \abovedisplayshortskip 0pt plus 3pt
  \belowdisplayshortskip 7pt plus 3pt minus 4pt
  \smallskipamount=3pt plus1pt minus1pt
  \medskipamount=6pt plus2pt minus2pt
  \bigskipamount=12pt plus4pt minus4pt
  \def\rm{\fam0\tenrm}          \def\it{\fam\itfam\tenit}%
  \def\sl{\fam\slfam\tensl}     \def\bf{\fam\bffam\tenbf}%
  \def\smc{\tensmc}             \def\mit{\fam 1}%
  \def\cal{\fam 2}%
  \textfont0=\tenrm   \scriptfont0=\sevenrm   \scriptscriptfont0=\fiverm
  \textfont1=\teni    \scriptfont1=\seveni    \scriptscriptfont1=\fivei
  \textfont2=\tensy   \scriptfont2=\sevensy   \scriptscriptfont2=\fivesy
  \textfont3=\tenex   \scriptfont3=\tenex     \scriptscriptfont3=\tenex
  \textfont\itfam=\tenit
  \textfont\slfam=\tensl
  \textfont\bffam=\tenbf \scriptfont\bffam=\sevenbf
  \scriptscriptfont\bffam=\fivebf
  \normalbaselines\rm}
\font\fourteenrm=cmr10 scaled 1440    \font\fourteeni=cmmi10 scaled 1440
\font\fourteensy=cmsy10 scaled 1440   \font\fourteenex=cmex10 scaled 1440
\font\fourteenbf=cmbx10 scaled 1440   \font\fourteensl=cmsl10 scaled 1440
\font\fourteentt=cmtt10 scaled 1440   \font\fourteenit=cmti10 scaled 1440
\font\fourteensc=cmcsc10 scaled 1440
\skewchar\fourteeni='177   \skewchar\fourteensy='60
\def\fourteenpoint{\normalbaselineskip=16pt
  \abovedisplayskip 14.8pt plus 4pt minus 10pt
  \belowdisplayskip 14.8pt plus 4pt minus 10pt
  \abovedisplayshortskip 0pt plus 4pt
  \belowdisplayshortskip 9.2pt plus 4pt minus 6pt
  \smallskipamount=4.4pt plus1.8pt minus1.8pt
  \medskipamount=8.8pt plus2.8pt minus2.8pt
  \bigskipamount=17.3pt plus5.6pt minus5.6pt
  \def\rm{\fam0\fourteenrm}          \def\it{\fam\itfam\fourteenit}%
  \def\sl{\fam\slfam\fourteensl}     \def\bf{\fam\bffam\fourteenbf}%
  \def\mit{\fam 1}                 \def\cal{\fam 2}%
  \def\sc{\fourteensc}               \def\tt{\fourteentt}
  \def\sf{\fourteensf}
  \textfont0=\fourteenrm   \scriptfont0=\tenrm   \scriptscriptfont0=\sevenrm
  \textfont1=\fourteeni    \scriptfont1=\teni    \scriptscriptfont1=\seveni
  \textfont2=\fourteensy   \scriptfont2=\tensy   \scriptscriptfont2=\sevensy
  \textfont3=\fourteenex   \scriptfont3=\fourteenex
                           \scriptscriptfont3=\fourteenex
  \textfont\itfam=\fourteenit
  \textfont\slfam=\fourteensl
  \textfont\bffam=\fourteenbf \scriptfont\bffam=\tenbf
                                            \scriptscriptfont\bffam=\sevenbf
  \normalbaselines\rm}
\font\twelverm=cmr10 scaled 1200    \font\twelvei=cmmi10 scaled 1200
\font\twelvesy=cmsy10 scaled 1200   \font\twelveex=cmex10 scaled 1200
\font\twelvebf=cmbx10 scaled 1200   \font\twelvesl=cmsl10 scaled 1200
\font\twelvett=cmtt10 scaled 1200   \font\twelveit=cmti10 scaled 1200
\font\twelvesc=cmcsc10 scaled 1200
\skewchar\twelvei='177   \skewchar\twelvesy='60


\def\twelvepoint{\normalbaselineskip=12.4pt plus 0.1pt minus 0.1pt
  \abovedisplayskip 12.4pt plus 3pt minus 9pt
  \belowdisplayskip 12.4pt plus 3pt minus 9pt
  \abovedisplayshortskip 0pt plus 3pt
  \belowdisplayshortskip 7.2pt plus 3pt minus 4pt
  \smallskipamount=3.6pt plus1.2pt minus1.2pt
  \medskipamount=7.2pt plus2.4pt minus2.4pt
  \bigskipamount=14.4pt plus4.8pt minus4.8pt
  \def\rm{\fam0\twelverm}          \def\it{\fam\itfam\twelveit}%
  \def\sl{\fam\slfam\twelvesl}     \def\bf{\fam\bffam\twelvebf}%
  \def\mit{\fam 1}                 \def\cal{\fam 2}%
  \def\sc{\twelvesc}               \def\tt{\twelvett}
  \def\sf{\twelvesf}
  \textfont0=\twelverm   \scriptfont0=\tenrm   \scriptscriptfont0=\sevenrm
  \textfont1=\twelvei    \scriptfont1=\teni    \scriptscriptfont1=\seveni
  \textfont2=\twelvesy   \scriptfont2=\tensy   \scriptscriptfont2=\sevensy
  \textfont3=\twelveex   \scriptfont3=\twelveex  \scriptscriptfont3=\twelveex
  \textfont\itfam=\twelveit
  \textfont\slfam=\twelvesl
  \textfont\bffam=\twelvebf \scriptfont\bffam=\tenbf
  \scriptscriptfont\bffam=\sevenbf
  \normalbaselines\rm}

\hoffset=2truein
\hsize=4.5truein
\def\singlespace{\baselineskip=\normalbaselineskip}
\parindent=40pt
\newcount\firstpageno
\firstpageno=2
\footline={\ifnum\pageno<\firstpageno{\hfil}\else{
 \hskip 3.19truein\twelverm\folio\hfill}\fi}
\def\gtwid{\mathrel{\raise.3ex\hbox{$>$\kern-.75em\lower1ex\hbox{$\sim$}}}}
\def\ltwid{\mathrel{\raise.3ex\hbox{$<$\kern-.75em\lower1ex\hbox{$\sim$}}}}
\rightline{UFIFT-HEP-95-5}
\vskip1truein
\fourteenpoint
\centerline{\bf Sources and Distributions of Dark Matter*}

\vskip .2truein

\centerline{Pierre Sikivie}
\vskip .2truein
\tenpoint
\centerline{\it I. Newton Institute, University of Cambridge, Cambridge, CB3
OEH,
UK}
\centerline{\it and}
\centerline{\it Physics Department, University of Florida, Gainesville, FL
32611,
USA}
\vskip 1truein

\noindent In the first section, I will try to convey a sense of the variety
of observational inputs that tell us about the existence and the
spatial distribution of dark matter in the universe.  In the second
section, I will briefly review the four main dark matter
candidates, taking note of each candidate's status in the world of
particle physics, its production in the early universe, its effect
upon large scale structure formation and the means by
which it may be detected.
Section 3 concerns the energy spectrum of (cold) dark matter
particles on earth as may be observed some day in a direct
detection experiment.  It is a brief account of work done in
collaboration with J.~Ipser and, more recently, with I.~Tkachev and
Y.~Wang.
\vskip 2.5truein
\vbox{\hsize=2.0in\hbox to \hsize{\leaders\hrule\hfil}}
{\tenpoint \noindent\singlespace
*To appear in the Proceedings of the Conference ``Trends in Astroparticle
Physics'',
Stockholm, Sweden, Sept. 22--25, 1994, Nucl. Phys. B. Proc. Supplements, edited
by L. Bergstrom, P. Carlson, P.O. Hulth and H. Snellman.}
\vfill\eject
\hsize=6.3truein
\hoffset=0.1truein
\vsize=8.5truein
\voffset=0truein
\topskip=0truein
\parskip=\medskipamount
\twelvepoint
\singlespace

Ref. [1] is a list of works which I have
consulted in preparing this rather cursory overview,  and which the
reader should turn to for more complete and in-depth information.
\vskip .15truein
\noindent{\bf I. DARK MATTER OBSERVATIONS}
\vskip .1truein
In 1932, Oort$^{[2]}$ studied the motion of galactic disk stars in
the vertical direction, i.e., perpendicular to the disk.  By
applying a version of the virial theorem to the distribution of
vertical star velocities, he obtained an estimate of the density of
the galactic disk in the solar neighborhood:
$$\rho_{\rm disk} \simeq 1.2 \cdot 10^{-23} {\rm gr}/{\rm cm}^3\,
.\eqno(1.1)$$
On the other hand, if one adds up the densities of all the matter
``seen'' in stars and interstellar gas, plus what is expected from
stellar remnants, mainly white dwarfs, one finds considerably less
than the dynamical estimate of Eq.~(1.1), of order half thereof.  So
there is dark matter in the galactic disk.
Because this dark matter is in the disk rather than the halo, we
expect it to be dissipative, which means in all probability that it
is baryonic dark matter.

In 1933, Zwicky$^{[3]}$ used measurements of the line-of-sight
velocities of galaxies in the Coma cluster to estimate the mass of
that cluster using the virial theorem.  The result he obtained in
this way is approximately 400 times the mass inferred by counting
the number of galaxies in the cluster and assigning to each a mass
($\sim 10^{11}M_\odot$) typical for the luminous part of a spiral
galaxy.  The masses of the luminous parts of some spiral galaxies
had already been determined by measuring their rotation curves up to
distances from their centers of order their disk radii.
Smith$^{[4]}$ obtained a similar result for the Virgo cluster.

In 1973, Ostriker and Peebles$^{[5]}$ pointed out that the tendency
of galactic disks to be unstable towards a large-scale bar mode can
be cured by assuming the existence of a spherical halo of dark
matter with mass within the disk radius of order the disk mass
($6\cdot 10^{10}M_\odot$ for our galaxy).  Thus the galactic mass
within a sphere with radius equal to the disk radius would be
roughly half in the disk and half in an unseen spherical halo.  Of
course, by Birkhoff's theorem,  the argument does not say anything
about halo matter outside the disk radius.

During the seventies, the rotation curves of spiral galaxies were
measured$^{[6]}$ over much larger distances than before,
in many cases extending the rotation curve to distances several
times the disk radius.  In all cases, the rotation velocity was
found to be constant (i.e., independent of radius $r$) or slightly
rising, up to the last measured point.  Balancing centrifugal and
gravitational forces, one has
$${GM(r)\over r^2} = {v_{rot}(r)^2\over r}\eqno(1.2)$$
where $M(r)$ is the mass interior to $r$ and $v_{rot}(r)$ is the
rotation velocity at $r$.    If only luminous matter were
contributing to the galactic mass, we would have $M(r)\sim$
constant and hence $v_{rot}(r)\sim r^{-{1\over 2}}$, for $r >$ disk
radius.  Instead, the data show $v_{rot}(r)\sim$ constant there and
hence $M(r)\sim r$.  The implication is that there is a halo of
dark matter whose density $\rho_{dm}(r)\sim {1\over r^2}$ at large
$r$.  The halo distribution is usually modeled by the function
$$\rho_{dm}(r) = {\rho_{dm} (0)\over 1+\left ({r\over a}\right )^2}
\eqno(1.3)$$
where $a$ is called the core radius.  For our own galaxy, $v_{rot}
\simeq 220$km/s, $a\simeq$ few kpc, and :
$$\rho_{dm} (r_\odot) \simeq {1\over 2} \cdot 10^{-24}{\rm gr}/{\rm
cm}^3\eqno(1.4)$$
where $r_\odot \simeq 8.5$~kpc is our distance to the galactic
center.  The estimate (1.4) of the local dark halo density is based
upon models$^{[7]}$ of the galactic mass distribution developed in
the early 80's by Bahcall and Soneira, and Caldwell and Ostriker.
However, the discovery of an abundance of microlensing events in
the direction of the galactic bulge has stimulated a lot of recent
work on the galactic mass distribution and this will likely result in
a more precise determination of the galactic halo  parameters.

In 1972, J. Einasto, A. Kaasik and E. Saar$^{[8]}$ studied 105 pairs of
galaxies, the members of each pair being close on the sky and assumed
to be gravitationally bound to each other.  They compared the
distance between each pair to its relative velocity to obtain an
estimate of its reduced inner mass.  Of course, only the
line-of-sight velocities and the angular projections of distance
onto the sky are measured and therefore an average over many pairs
must be performed to try and eliminate the effects of projection
and ignorance of orbit eccentricities.  At any rate, these authors
find that the galactic mass increases with distance, approximately
linearly, up to masses of order $10^{13}M_\odot$.  This study and
others$^{[1]}$ imply that galactic halos extend very far out.  I
do not know of anything that contradicts the assumption that
galactic halos extend all the way to radii of order 1--2
Mpc, where the halo density, falling off as $1\over r^2$, becomes
equal to the average intergalactic dark matter density.

There are a number of methods to estimate the average dark matter
density on scales larger than the typical intergalactic or
intercluster distance ($\gtwid 10$~Mpc).  Density perturbations on
such large scales are still in the linear regime of their growth by
gravitational instability.  Let us describe a particular method.
If a region has an overdensity $\delta\rho$ in excess of the
average density $\rho$, neighboring galaxies will have an excess
gravitational attraction towards that region and consequently
deviate from perfect Hubble flow.  One writes:
$$\vec v = H_0 \vec r + \vec v_p\eqno(1.5)$$
where $H_0$ is the Hubble expansion rate, $\vec r$ is the position
relative to the center of an overdensity and  $\vec v_p$ is
called the peculiar velocity.  It is found$^{[9]}$ that in
the linear regime around a single, spherically symmetric overdensity
$$\vec v_p =- H_0 \vec r {1\over 3} \Omega^{0.6} {\delta\rho\over
\rho}\, ,\eqno(1.6)$$
where
$$\Omega ={\rho\over \rho_{crit}} ={8\pi G\rho\over 3H_0^2}\, .
\eqno(1.7)$$
$\rho_{crit}$ is the critical density for closing the universe.
The $\Omega$ dependence on the RHS of Eq.~(1.6) is a close fit to
the actual $\Omega$ dependence for vanishing cosmological constant.
Eq.~(1.6) affords a way to determine $\Omega$ by measuring peculiar
velocities $\vec v_p$ and comparing them with observed
overdensities $\delta\rho\over \rho$.  However, $\delta \rho\over
\rho$ cannot be measured directly.  What can be done is count
galaxies, measuring their average density $n_G$ and local
overdensities $\delta n_G$.  Unfortunately, the relationship
between $\delta\rho\over \rho$ and $\delta n_G\over n_G$ is not
known.  It is parametrized by a fudge factor $b$, called
the ``bias parameter'':
$${\delta n_G\over n_G} =b{\delta \rho\over \rho}\, .\eqno(1.8)$$
So the method of peculiar velocities actually measures
${\Omega^{0.6}\over b}$.
A number of authors$^{[10]}$ have analyzed galaxy distributions
in this way with the result:
$${\Omega^{0.6}\over b}=1\pm 0.3\, .\eqno(1.9)$$
Most attempts to determine the bias parameter from first principles
yield $b>1$.  Eq.~(1.9) suggests then that, when measured on the
largest scales, the value of the density parameter $\Omega$ is
consistent with a critically closed universe ($\Omega=1$).  On the
other hand, the measurements on these large scales are very
imprecise.

Of course, $\Omega=1$ is strongly favored on `theoretical' grounds.
A $\Omega \neq 1$ universe will deviate from $\Omega=1$ more and
more as time goes on, in pretty much the same way as a pencil
standing nearly vertically on its point will fall over.  For a
universe to stay near $\Omega =1$ for a long time, it has
to be extraordinarily close to $\Omega=1$ to start with.  This
problem of initial conditions for our universe is called the
flatness (or age) problem.  It may be neatly solved by assuming
that there is, at very early times, a brief epoch during
which the energy density is dominated by vacuum energy density and,
as a result, the universe expands at an exponential rate.  After
this ``inflation'', $\Omega=1$ with tremendous precision.  The
inflationary cosmology has many other attractive features as
well.$^{[11]}$\ \ So there are compelling reasons to believe that
$\Omega=1$.  Whether observations support this prejudice
is not obvious, although it seems fair to say that they are in
rough agreement with it.  Luminous matter contributes $\Omega_{lum}
\simeq 3.10^{-3}$ to $6.10^{-3}$.  Dark matter in galactic halos and
in clusters of galaxies contributes $\Omega_{gal} \simeq 0.02$ to
$0.2$.  Finally, as Eq.~(1.9) and the results of other observations
on the largest scales studied suggest, there may be enough dark
matter not associated with galaxies or clusters of galaxies (in
voids, say) to yield $\Omega =1$.

The success of nucleosynthesis$^{[12]}$ in producing the primordial
abundances of light elements requires that the contribution
$\Omega_B$ of baryons satisfies:
$$0.011\leq 0.011 h^{-2} \leq \Omega_B \leq 0.019 h^{-2}\leq
0.12\eqno(1.10)$$
where $h$ parametrizes the present Hubble expansion rate
$$H_0= 100 h \cdot {km\over s\ Mpc}\, .\eqno(1.11)$$
Measurements of $H_0$ are in the range of $0.4\leq h \leq 1$.
Since $\Omega_{lum} < 0.006$, Eq.~(1.10) implies that some baryons
are dark.  Recall that there is dark matter associated with the
disk of our galaxy.  Because it is in the disk rather than in a
halo, this dark matter must be dissipative which presumably means
that it is baryonic.  (I am assuming that we have necessarily
discovered the existence of any form of matter sufficiently
abundant and sufficiently strongly interacting to be the disk dark
matter.   Disk matter must have sufficiently strong
interactions to have concentrated in a disk by dissipating its
energy while conserving its angular momentum.)  Moreover, the
recent discovery$^{[13]}$  of microlensing in the direction of the galactic
bulge indicates that our disk has a population of low mass ($\sim
0.1 M_\odot$) compact objects.  These may be `brown dwarfs', i.e.
stars too low in mass to shine by nuclear burning, which are the most
likely hiding place for dark baryons.$^{[14]}$

Eq.~(1.10) is compatible with the assumption that all the dark
matter in galaxies and clusters of galaxies is in baryons.
However, we will see in the next section that this economical
hypothesis runs into difficulties in most scenarios of galaxy
formation.  Eq.~(1.10) is not compatible with $\Omega_B=1$.\ \
Hence, if one believes in inflation and in standard
nucleosynthesis---and both of these are very well motivated---one
must conclude that our universe is dominated, at the
90\% level, by a form of dark matter which is not baryonic.
\vskip .2truein
\noindent{\bf 2. DARK MATTER CANDIDATES}
\vskip .1truein
\noindent{\bf 2.1.  Baryons}

They are known to exist.  Moreover, they are known to be a
form of dark matter, and the nucleosynthesis constraints allow all
dark matter associated with galaxies to be
in baryons.  Hence, a conservative hypothesis may be that
$\Omega\simeq \Omega_B \simeq 0.2$.\ \  (Inflation is given up
then.)\ \ However, that particular scenario has serious
difficulties with galaxy formation.  The point is that the
density perturbations in the baryon distribution that should
produce galaxies cannot grow by gravitational instability till
after the epoch of recombination at a temperature $T_{rec} \simeq
4.10^3$~K.  Recombination is when electrons combine with ions to
form neutral atoms.  Before recombination, the baryons are
in close thermal contact with the photon gas.  Because the latter
has pressure, the Jean's mass is large:
$$M_J\sim 1.8\cdot 10^{16}M_\odot (\Omega_B h^2)^{-2} \quad
\hbox{ \ for}\quad  T>T_{rec}\eqno(2.1)$$
in this scenario.  The Jean's mass sets the critical scale below
which density perturbations do not grow.  After recombination,
$M_J\sim 0$ and density perturbations in the matter
distribution grow on all scales at the rate ${\delta\rho\over
\rho}\sim R$ where $R$ is the cosmological scale factor.
 The temperature drops as $T\sim R^{-1}$.\ \ Galaxies form when
${\delta\rho\over \rho} \sim 1$ on the appropriate mass scale,
about $10^{12}M_\odot$.\ \ For this to happen before the
present, the density perturbations in the baryon distribution at
recombination must therefore have a minimum amplitude:
$$\left.{\delta\rho\over \rho}\right|_{rec} > {R_{rec}\over R_0} =
{T_0\over T_{rec}} = {2.73 K\over 4.10^3K} = 0.7\ 10^{-3}.\eqno(2.2)$$
One should expect accompanying photon temperature fluctuations of
the same order of magnitude.  These would contradict the upper
limits $\left ({\delta T\over T}\ltwid 10^{-5}\right )$ on the
microwave background anisotropy.  The scenario has additional
difficulties due to the diffusion of photons, which tends to erase
adiabatic fluctuations in the baryon number density.$^{[15]}$

The above model ($\Omega\simeq \Omega_B \simeq 0.2$) came first
historically and, as we just saw, it led to the expectation of
large CMBR anisotropies (${\delta T\over T} \simeq 10^{-3}$ or
$10^{-4}$) which got into more and more severe disagreement with
the observations.  This outcome provided a strong impetus for
the development of models with cold dark matter (CDM).  Indeed, the
CDM candidates decouple from photons and baryons.   As a result,
the density perturbations in CDM start to grow by gravitational
instability as soon as $t>t_{eq}$ where $t_{eq}$ is the
time of  equality of the radiation and matter energy densities
($\rho_{rad.}\sim
R^{-4}$, $\rho_{matt.}\sim R^{-3}$, $\rho_{rad.}=\rho_{matt.}$ at
$t_{eq}$).\ \   In models where $\Omega_{CDM}$ is close to one,
$t_{eq}$ comes well before $t_{rec}$.  Because there is more time
for their growth, the primordial density perturbations in these
models are smaller than the RHS of Eq.~(2.2).

At any rate, as we saw, some baryonic dark matter is known to exist
and there may be large amounts of it, up to $\Omega_B\simeq 0.2$.\
\ As already mentioned, a likely hiding place$^{[14]}$ for these dark
baryons is ``brown dwarfs'',
i.e., stars too low in mass to burn by nuclear fusion.  Paczynski$^{[16]}$
pointed out that objects of this kind, generically called MACHOs
for massive compact halo objects, can be searched for by looking
for the gravitational lensing of background stars by MACHOs that
happen to pass close to the line of sight.  Three collaborations$^{[13]}$
have reported compelling candidates for such microlensing events.
This very exciting development is reviewed by Ansari$^{[17]}$ at
this meeting.
\vskip .15truein
\noindent{\bf 2.2.  Neutrinos}

Neutrinos decouple in the early universe at a temperature
$T_D$ of order a few MeV.\ \ After that, each neutrino moves freely
and hence its momentum decreases with the universe's expansion
according to:  $p_\nu\sim R^{-1}$.  Thus, neglecting
inhomogeneities, the neutrino phase-space density is given by:
$${\cal N}\; (\vec r,\vec p) ={g_\nu\over (2\pi)^3} {1\over e^{p\over
T_\nu(t)}+1}\, ,\eqno(2.3)$$
where $T_\nu(t)\equiv T_D{R_D\over R(t)}$ and where $g_\nu$ is the
number of neutrino spin  degrees of freedom.  In the standard
model, each neutrino flavor contributes $g_\nu=2$.  $T_\nu(t)$ is
usually called the `neutrino temperature' although the
distribution (2.3) will deviate from a thermal one if the neutrino
is massive.  The photon temperature also decreases according to
$T_\gamma \sim R^{-1}$ most of the time.  If this were always true,
$T_\gamma$ and $T_\nu$ would remain equal.  However, at a
temperature of order 1~MeV, electrons and positrons annihilate,
thus reheating the photon gas.  Because $e^+e^-$ annihilation is
adiabatic (the process goes back and forth very rapidly compared to
the Hubble expansion rate), conservation of entropy allows one to
relate the photon temperature after annihilation to the temperature
the photons would have had if there had been no annihilation, which
is the neutrino temperature.  This yields the famous result:
$$T_\nu=\left({4\over 11}\right)^{1\over 3} T_\gamma\eqno(2.4)$$
after $e^+e^-$ annihilation.  Since $T_\gamma=2.73$~K today,
$T_\nu=1.95$~K.  This implies in particular that the number density
of neutrinos and anti-neutrinos today $n_{\nu+\overline\nu}
(t_0)=113/{\rm cm}^3$ per neutrino flavor.  From this
one readily finds that, in extensions of the standard model where
the neutrinos have small Majorana masses, their contribution to the
cosmological energy density is$^{[18]}$
$$\Omega_\nu h^2 = \sum_i {m_{\nu i}\over 94 eV}\, ,\eqno(2.5)$$
where the sum is over flavors. Although none of these neutrinos
have been observed directly or indirectly, we are confident that they
are there because the theoretical arguments for their existence are
very simple and conservative.

So the next assumption we will consider is that neutrinos
constitute most of the dark matter and hence that they dominate the
cosmological energy density.  This scenario also tends to run into
trouble with galaxy formation.  As was already mentioned,
perturbations in the matter density only start to grow after
matter-radiation equality.  In the present scenario, equality
occurs when the temperature is of order the neutrino mass.  Before
that the neutrinos are relativistic and their ``free-stream\-ing''
erases all density perturbations in the neutrino fluid on length
scales less than the free-stream\-ing distance,$^{[19]}$ i.e., the
distance a typical neutrino travels from the Big Bang till the time
of equality.  The corresponding mass scale is:
$$M_\nu=4\cdot 10^{15} M_\odot \left ({30 eV\over m_\nu}\right)^2\,
,\eqno(2.6)$$
which is of order the mass in large galactic clusters.  The
resulting spectrum of primordial density perturbations is heavily
suppressed on all mass scales less than $M_\nu$, including the mass
scale ($\sim 10^{12}M_\odot$) of individual galaxies.  If such a
spectrum is used as input in computer simulations of large scale
structure formation, a poor fit to the observations
results.$^{[20]}$\ \ The difficulties neutrinos have with large
scale structure formation may be eased if topological
defects,$^{[21]}$ such as cosmic strings, are the source of the
density perturbations because in this case, the density
perturbations continue to be created after $t_{eq}$.

There is however a separate difficulty with neutrinos constituting
galactic halos.$^{[22]}$\ \ Liouville's theorem tells us that the
phase-space density is constant following the motion.  Eq.~(2.3)
therefore implies that the neutrino phase-space density can nowhere
be larger than ${1\over 2}g_\nu$.  On the other hand, for neutrinos
to constitute a galactic halo, their velocities must be less
than the escape velocity, which for our own galaxy is of order
$10^{-3}c$.  The upper limits on the phase-space density and on the
velocity imply the following upper limit on the physical density:
$$\rho_{\nu, max} ={1\over 2} {10^{-24}gr\over cm^3}
\left({v_{max}\over 10^{-3}c}\right)^3 \left({m_\nu\over
19eV}\right)^4\, .\eqno(2.7)$$
Eq. (1.3), (1.4), (2.5) and (2.7) tell us that neutrinos can only barely
be packed tightly enough to constitute the Milky Way halo.  Dwarf
galaxies also have dark matter halos but smaller escape velocities.
For these galaxies, the neutrino phase-space constraint is severely
violated.$^{[23]}$
\vskip .15truein
\noindent{\bf 2.3.  WIMPs}

WIMPs is an acronym for Weakly Interacting Massive
Particles.  To be specific, consider a massive neutral lepton $L$.\
\ If its mass $m_L$ is less than the temperature $T_D$ at which it
decouples from the thermal bath, then $L$ behaves like a neutrino
and its cosmological energy density is given by Eq.~(2.5) or
something very similar to it.  However,$^{[24]}$ if $m_L$ exceeds $T_D$,
then for $m_L > T > T_D$, the number density of $L$ particles:
$$n_L(T)=n_{eq}(T) = {g_L\over (2\pi)^3} \int_0^\infty 4\pi p^2dp{1\over
e^{\sqrt{m_L^2+p^2}\over T}+1}\eqno(2.8)$$
falls off exponentially, as $e^{-m_L\over T}$.\ \ The number
density $n_L(T)$
tracks its equilibrium value  $n_{eq}(T)$ as long as the
annihilation rate of $L$ particles exceeds the Hubble rate.  Thus
$T_D$ is given by:
$$\langle \sigma_{ann}v\rangle n_{eq}(T_D)\simeq H(T_D)\eqno(2.9)$$
where $\sigma_{ann}$ is the annihilation cross-section of $L$
particles.  In Eq.~(2.9), the dominant dependence upon $T_D$ is the
exponential $e^{-m_L\over T_D}$ behaviour of $n_{eq}(T_D)$.  As a
consequence, $T_D$ is proportional to $m_L$ up to logarithmic
corrections.  For cross-sections typical of weakly interacting
particles, one finds:
$$T_D\simeq {1\over 20} m_L\, .\eqno(2.10)$$
Hence, the cosmological energy density in $L$ particles today:
$$\rho_L(t_0) = m_L n_L(t_0) = m_L n_L (t_D)
\left({R_D\over R_0}\right)^3
\simeq m_L {H(T_D)\over \langle \sigma_{ann} v\rangle} {N_0\over N_D}
\left ({T_0\over T_D}\right)^3\eqno(2.11)$$
where $N_0$ and $N_D$ are the effective numbers of thermal degrees
of freedom today and at the decoupling of $L$ particles,
conservation of entropy from $t_D$ till $t_0$ having been assumed.
Remarkably, the $T_D$ dependence on the RHS of Eq.~(2.11) cancels
out because of Eq.~(2.10) and because $H(T_D)=\sqrt{{8\pi G\over
3}\rho(T_D)} = \sqrt{{8\pi G\over 3} N_D {\pi^2\over 30} T_D^4}$.\
\ As a result, the contribution of $L$ particles, and more
generally that of WIMPs, to the cosmological energy density depends
almost exclusively upon their annihilation cross-section.  One
finds:
$$\Omega_{\rm WIMP} h^2\simeq {6\cdot 10^{-27}\over \langle
\sigma_{ann} v\rangle} {{\rm cm}^3\over{\rm sec}}\, .\eqno(2.12)$$
For the particular case of a heavy neutral lepton $L$, one has
$\sigma_{ann} \sim G_F^2m_L^2$ for $m_L\ltwid m_Z=91.2$~GeV and
$\sigma_{ann}\sim {\alpha^2\over m_L^2}$ for $m_L\gtwid m_Z$.  In
that case, $\Omega_L h^2 \sim {1\over m_L^2}$ in the range few
MeV~$<m_L<m_Z$ with $\Omega_L h^2=1$ for $m_L\simeq 2$~GeV,
and $\Omega_L h^2\sim m_L^2$ when $m_L\gtwid m_Z$ with $\Omega_L
h^2 =1$ for $m_L\simeq 10$~TeV.\ \ Note that if the WIMP is not its
own anti-particle, there may be a WIMP-antiWIMP asymmetry, similar
to the baryon asymmetry.  In that case there is an additional
contribution to $\Omega_{\rm WIMP} h^2$, aside from the
one given by Eq.~(2.12).

The best motivated WIMP candidate is the lightest supersymmetric
partner (LSP) in supersymmetric extensions of the standard model.
Typically, this is a linear combination of the pho\-tino, the zino
and the Higgsino.

Because WIMPs are non-relativistic from the moment of their
decoupling, their free-stream\-ing distance is very small and hence
their free-stream\-ing does not erase density perturbations on any
relevant scales.  For this reason, WIMPs are called Cold Dark
Matter (CDM).  In contrast, neutrinos, because of their large
free-stream\-ing distance, are called Hot Dark Matter (HDM).  The
assumptions of CDM, with $\Omega_{CDM}\simeq 1$, and of a flat
(Zel'dovich-Harrison) spectrum of primordial density perturbations
yield a model of large scale structure formation$^{[25]}$ which has been
thoroughly tested and which has been, by and large, very
successful.  However, in light of the COBE measurement
of the cosmic microwave anisotropy which is also a measurement
of the primordial density perturbations on the largest scales observable,
some modification of the pure CDM model may be required.

If WIMPs constitute the halo of our galaxy, they may be searched
for on earth by looking for WIMP + nucleus elastic scattering in a
laboratory detector.$^{[26]}$  The nuclear recoil can be put into
evidence by low temperature calorimetry, by ionization detection or
by the detection of ballistic phonons.$^{[27]}$  WIMPs may also be
searched for by looking for the decay products (photons,
anti-protons$\ldots$) of WIMP annihilation in the halo of our
galaxy$^{[28]}$ or by looking for neutrinos produced by the
annihilation of WIMPs that have been captured by the sun.$^{[29]}$
\vskip .15truein
\noindent{\bf 2.4.  Axions}

The axion is a hypothetical particle whose existence would
insure that the strong interactions conserve $P$ and $CP$ in spite
of the fact that other interactions violate those symmetries.$^{[30]}$
Indeed, the action density of the standard model of elementary particles
contains in general a term:
$$L_\theta = {\theta g_s^2\over 32\pi^2} G_{\mu\nu}^a \tilde G^{a\,
\mu\nu}\eqno(2.13)$$
where $G_{\mu\nu}^a$ is the gluonic field strength, $\tilde
G_{\mu\nu}^a$ is the dual of $G_{\mu\nu}^a$, and $g_s$ is the QCD
gauge coupling.  If $\theta\neq 0$, non-perturbative QCD effects
induce violations of $P$ and $CP$ in the strong interactions.  No
such violation has been observed.  In particular, the upper limit
on the neutron electric dipole moment requires $\theta< 10^{-9}$.
But there is no reason in the standard model for the parameter
$\theta$ to be small.  This shortcoming has been called the
``strong CP problem''.

Peccei and Quinn modified the standard model in such a way that the
parameter $\theta$ in Eq.~(2.13) gets replaced by ${a(x)\over f_a}$
where $a(x)$ is a dynamical pseudo-scalar field whose quantum is
called the axion ; $f_a$ is a quantity with dimension of energy
called the axion decay constant.  By construction, the vacuum
expectation value of $a(x)$ is indifferent except for those
non-perturbative effects that make QCD depend upon $\theta$.  The
latter produce an effective potential $V(\theta) =
V\left({a(x)\over f_a}\right)$ whose minimum is at $\theta=0$.
Thus by postulating an axion, $\theta$ is allowed to relax to zero
dynamically and the strong $CP$ problem is solved.

The properties of the axion can be derived using the methods of
current algebra.  The axion mass is related to $f_a$ by:
$$m_a \simeq 0.6 {\rm eV} {10^7{\rm GeV}\over f_a}\, .\eqno(2.14)$$
All the axion couplings are inversely proportional to $f_a$.  Thus,
a very light axion is also very weakly coupled.$^{[31]}$  A priori,
the value of $f_a$, and hence that of $m_a$, is arbitrary.
However, astrophysical considerations$^{[32]}$ and searches for the
axion in high-energy and nuclear physics experiments$^{[33]}$ rule out
$m_a>10^{-3}$~eV.  On the other hand, cosmology places {\it a lower
limit} on $m_a$ of order $10^{-6}$~eV by requiring that axions do
not overclose the universe.$^{[34]}$

Indeed, for small masses, axion production in the early universe is
dominated by a novel mechanism.  The point is that the
non-perturbative QCD effects that produce the effective potential
$V\left({a(x)\over fa}\right)$ are strongly suppressed at
temperatures high compared to $\wedge_{\rm QCD}$.\ \ At these high
temperatures, $\langle a(x)\rangle$ has arbitrary value.  At
$T\simeq 1$~GeV, the potential $V$ turns on and the axion field
starts to oscillate about its CP conserving minimum.  These
oscillations do not dissipate in other forms of energy
because, in the relevant mass range, the axion is too weakly
coupled for that to happen.  The oscillations of the axion field
may be described as a fluid of axions.  The typical momentum of
these axions is the inverse of the correlation length of the axion
field at $T\simeq 1$~GeV.\ \ Since that correlation length
is of order the horizon then, we have $p_a\sim {1\over t_{1{\rm
GeV}}}\sim {1\over 10^{-6}{\rm sec}}\sim 10^{-9}$~eV.\ \ Thus, the
axion fluid is very cold compared to the ambient temperature.  Its
contribution to the present cosmological  energy density is found
to be of order
$$\Omega_a h^2 \simeq 0.3 \left ({10^{-6}{\rm eV} \over
m_a}\right)^{7\over 6} \left ({200 {\rm MeV}\over \wedge_{\rm
QCD}}\right)^{3\over 4}\, .\eqno(2.15)$$
Several sources of uncertainty affect the relationship
between $\Omega_a h^2$ and $m_a$, amongst which are the nature of
the QCD phase transition and the contribution to $\Omega_a h^2$
from cosmic axion strings.$^{[35,36]}$  Also, if inflation
occurs and the
post-inflation reheating temperature is less than $f_a$, then the
axion field gets homogenized and there may be an accidental
suppression of $\Omega_ah^2$ because the axion field happens to lie
everywhere close to the CP conserving minimum of $V$.\ \ From the
point of view of large scale structure formation, axions are cold
dark matter since they are non-relativistic from the moment of
their production during the QCD phase transition, as was emphasized
above.

Axion dark matter may be searched for by stimulating the conversion
of galactic halo axions to photons in a laboratory magnetic field.$^{[37]}$
The relevant coupling is
$$L_{a\gamma \gamma} =g_\gamma {\alpha\over \pi} {a\over fa} \vec E
\cdot \vec B\eqno(2.16)$$
where $g_\gamma$ is a model-dependent coupling constant of order
one.  If an electromagnetic cavity is permeated by a static
approximately homogeneous magnetic field $\vec B_0$ and the
resonant frequency of the lowest TM (relative to the direction of
$\vec B_0$) mode equals the axion mass, some galactic halo axions
will convert to quanta of that cavity mode.  For $B_0\sim 10$~Tesla
and cavity volumes of order 1m$^3$, the power from these $\alpha\to
\gamma$ conversions becomes detectable in a sufficiently short
amount of time to allow a search over a large axion mass range.
Two pilot experiments using this technique have been carried out.$^{[38]}$
At present, a second generation experiment$^{[39]}$ is under construction at
Lawrence Livermore National Laboratory that will be able to detect
dark matter axions if their local density is equal to the
local halo density given in Eq.~(1.4) or higher, if $g_\gamma \geq
1$ and if $m_a$ is in the range $1.3\mu$eV$\leq m_a\leq 13\mu$eV.
\vskip .2truein
\vbox{{\parindent=26pt
\noindent{\bf 3. THE PHASE-SPACE STRUCTURE OF COLD DARK MATTER
\break\indent
HALOS}}
\vskip .1truein
If a signal is found in the cavity detector of galactic halo
axions, it will be possible to measure the energy spectrum with
great precision and resolution because all the time that was
previously used in searching for the signal can now be used to
accumulate data.  Hence there is good motivation to ask what can be
learned about our galaxy from analyzing such a signal. }

In many past discussions of dark matter detection on earth, it has
been assumed that the dark matter  particles have an isothermal
distribution.  Thermalization has been argued to be the result of
a period of ``violent relaxation''$^{[40]}$ following the collapse of the
protogalaxy.  If it is strictly true that the velocity distribution
of dark matter particles is isothermal, which seems to be a
strong assumption, then the only information that can be gained
from its observation is the corresponding virial velocity and our
own velocity relative to its standard of rest.  If, on the other
hand, the thermalization is incomplete, a signal in a dark matter
detector may yield additional information.

J.R. Ipser and I discussed$^{[41]}$ the extent to which the phase-space
distribution of cold dark matter particles is thermalized in a
galactic halo and concluded that there are substantial deviations
from a thermal distribution in that the highest energy particles
have discrete values of velocity.  There is one velocity peak on
earth due to dark matter particles falling onto the galaxy for the
first time, one peak due to particles falling out of the galaxy for
the first time, one peak due to particles falling into the galaxy
for the second time, etc.  The peaks due to particles that have
fallen in and out of the galaxy a large number of times in the past
are washed out because of scattering in the gravitational wells of
stars, globular clusters and large molecular clouds.  But the peaks
due to particles which have fallen in and out of the galaxy only a
small number of times in the past are not washed out.

If the fraction of the local dark matter density which is in these
velocity peaks is sufficiently large, a direct dark matter search,
such as the LLNL experiment, may be made more sensitive by having
it look specifically for velocity peaks.  I.~Tkachev, Y.~Wang and
I have been studying galactic halo formation with the purpose of
obtaining estimates of the sizes and locations of the velocity
peaks.$^{[42]}$  To this end, we have generalized the secondary
infall model of
galactic halo formation to include angular momentum of the dark
matter particles.  This new model is still spherically symmetric
and it has self-similar solutions.  We find that the typical
fraction of the local cold dark matter density in any one of the
highest energy velocity peaks is several percent.  A forthcoming
paper will give estimates of the highest energy
peaks as a function of the amount of angular momentum and other
model parameters.
\vskip .2truein
\noindent{\bf References}
\vskip .1truein
\item{1.} J. Binney and S. Tremaine, {\it Galactic Dynamics},
Princeton U. Press,
1987;\ \  E.W. Kolb and M.S. Turner, {\it The Early Universe},
Addison-Wesley,
1988;\ \ P.J.E. Peebles, {\it Principles of Physical Cosmology},
Princeton U.
Press, 1993;\ \ V. Trimble, Ann. Rev. Astron. Astroph. {\bf 25}
(1987) 425;\ \
C. Jones and A. Melissinos, editors, {\it Cosmic Axions}, World
Scientific, 1989;\ \
M. Srednicki, Editor,
{\it Particle Physics and Cosmology:  Dark Matter}, North-Holland., 1990;
\ \ M. Turner, Phys. Scripta {\bf T36} (1991) 167.
\item{2.} J. Oort, Bull. Astr. Inst. Netherlands {\bf 6} (1932)
249.
\item{3.} F. Zwicky, Helv. Phys. Acta {\bf 6} (1933) 110.
\item{4.} S. Smith, Ap. J. {\bf 83} (1936) 23.
\item{5.} J.P. Ostriker and P.J.E. Peebles, Ap. J. {\bf 186} (1973)
467.
\item{6.} V.C. Rubin and W.K. Ford, Ap. J. {\bf 159} (1970) 379;
D.H. Rogstad and
G.S. Shostak, Ap. J. {\bf 176} (1972) 315; V.C. Rubin, W.K. Ford
and N. Thonnard,
Ap. J. {\bf 238} (1980) 471.
\item{7.} J.N. Bahcall and R.M. Soneira, Ap. J. Suppl. {\bf 44}
(1980) 73; J.A.R.
Caldwell and J.P. Ostriker, Ap.J. {\bf 251} (1981) 61.
\item{8.} J. Einasto, A. Kaasik and E. Saar, Nature {\bf 250} (1974) 309.
\item{9.} P.J.E. Peebles, {\it Principles of Physical Cosmology},
Princeton U. Press, 1993.
\item{10.} A. Dressler et al., Ap. J. {\bf 313} (1987) L37;
E. Bertschinger and A. Dekel, Ap. J.  {\bf 336} (1989) L5;
M. Rowan-Robinson et al., MNRAS {\bf 247} (1990) 1.
\item{11.} For a review, see for example:  {\it Inflationary
Cosmology}, edited by L. Abbott and S.-Y. Pi, World Scientifc, 1986.
\item{12.} See the review by L. Krauss in these proceedings.
\item{13,} C. Alcock et al., Nature {\bf 365} (1993) 621;
F. Aubourg et al., Nature {\bf 365} (1993) 623;
A. Udalski et al., Acta Astr. {\bf 43} (1993) 289.
\item{14.} B. Carr, Ann. Rev. Astron. Astroph. {\bf 32} (1994) 531.
\item{15.} J. Silk, Ap. J. {\bf 211} (1977) 638;
M.J. Rees and J.P. Ostriker, MNRAS {\bf 179} (1977) 541.
\item{16.} B. Paczynski, Ap. J. {\bf 304} (1986) 1.
\item{17.} R. Ansari, these proceedings.
\item{18.} S.S. Gershtein and Y.B. Zel'dovich, JETP Lett. {\bf 4}
(1966) 120.
\item{19.} J.R. Bond, G. Efstathion and J. Silk, Phys. Rev. Lett.
{\bf 45} (1980)
1980.
\item{20.} S.D.M. White, C.S. Frenk and M. Davis, Ap. J. {\bf 274}
(1983) L1.
\item{21.} For a review, see A. Vilenkin and E.P.S. Shellard, {\it Cosmic
Strings and other Topological Defects}, Cambridge U. Press, 1994.
\item{22.} S. Tremaine and J.E. Gunn, Phys. Rev. Lett. {\bf 42}
(1979) 407.
\item{23.} M. Aaronson, Ap. J. {\bf 266} (1983) L11; D.N.C. Lin and S.M.
Faber, Ap.J. {\bf 266} (1983) L21.
\item{24.} P. Hut, Phys. Lett. {\bf B69} (1977) 85; B.W. Lee and S.
Weinberg, Phys. Rev. Lett. {\bf 39} (1977) 165; M.I. Vysotskii, A.D. Dolgov
and Y.B. Zel'dovich, JETP Lett. {\bf 26} (1977) 188.
\item{25.} P.J.E. Peebles, Ap.J. {\bf 263} (1982) L1;
J. Ipser and P. Sikivie, Phys. Rev. Lett. {\bf 50} (1983) 925;
G.R. Blumenthal, S.M. Faber, J.R. Primack and M.J. Rees, Nature {\bf 311}
(1984) 517;
M. Davis, G. Efstathiou, C.S. Frenk and S.D.M. White, Ap. J. {\bf 292}
(1985) 371;
S.D.M. White, C.S. Frenk, M. Davis and G. Efstathiou, Ap. J. {\bf 313}
(1987) 505.
\item{26.} M. Goodman and E. Witten, Phys. Rev. {\bf D31} (1985)
3059; I. Wasserman, Phys. Rev. {\bf D33} (1986) 2071;
 K. Griest, Phys. Rev. {\bf D38} (1988) 2357.
\item{27.} S.P. Ahlen et al., Phys. Lett. {\bf B195} (1987) 603;
D.O. Caldwell et al., Phys. Rev. Lett. {\bf 61} (1988) 510; P.F. Smith
and J.D. Lewin, Phys. Rep. {\bf 187} (1990) 203.
\item{28.} J. Silk and M. Srednicki, Phys. Rev. Lett. {\bf 53}
(1984) 624.
\item{29.} W.H. Press and D.N. Spergel, Ap. J. {\bf 296} (1985)
679; J. Ellis, R. Flores and S. Ritz, Phys. Lett. {\bf B198} (1987) 393.
\item{30.} R.D. Peccei and H. Quinn, Phys. Rev. Lett. {\bf 38} (1977) 1440
and Phys. Rev. {\bf D16} (1977) 1791;
S. Weinberg, Phys. Rev. Lett. {\bf 40} (1978) 223;
F. Wilczek, Phys. Rev. Lett. {\bf 40} (1978) 279.
\item{31.} J.E. Kim, Phys. Rev. Lett. {\bf 43} (1979) 103;
M.A. Shifman, A.I. Vainshtein and V.I. Zakharov, Nucl. Phys. {\bf B166}
(1980) 493;
M. Dine, W. Fischler and M. Srednicki, Phys. Lett. {\bf 104B} (1981) 199;
A.P. Zhitnitskii, Sov. J. Nucl. {\bf 31} (1980) 260.
\item{32.} For a review, see:  M.S. Turner, Phys. Rep. {\bf 197} (1990) 67;
G.G. Raffelt, Phys. Rep. {\bf 198} (1990) 1.
\item{33.} For a review, see:  J.E. Kim, Phys. Rep. {\bf 150} (1987) 1;
H.-Y. Cheng, Phys. Rep. {\bf 158} (1988) 1;
R.D. Peccei, in {\it CP-Violation}, edited by C. Jarlskog, World
Scientific, 1989.
\item{34.} L. Abbott and P. Sikivie, Phys. Lett. {\bf 120B} (1983) 133;
J. Preskill, M. Wise and F. Wilczek, Phys. Lett. {\bf 120B} (1983) 127;
M. Dine and W. Fischler, Phys. Lett. {\bf 120B} (1983) 137.
\item{35.} R. Davis, Phys. Rev. {\bf D32} (1985) 3172 and Phys. Lett.
{\bf 180B} (1986) 225; A. Vilenkin and T. Vachaspati, Phys. Rev.
{\bf D35} (1987) 167; R.L. Davis and E.P.S. Shellard, Nucl. Phys.
{\bf B324} (1989) 167; A. Dabholkar and J.M.  Quashnock, Nucl.
Phys. {\bf B333} (1990) 815; R.A. Battye and E.P.S. Shellard,
Phys. Rev. Lett. {\bf 73} (1994) 2954.
\item{36.} D. Harari and P. Sikivie, Phys. Lett. {\bf B195} (1987) 361;
C. Hagmann
and P. Sikivie, Nucl. Phys. {\bf B363} (1991) 247.
\item{37.}P. Sikivie, Phys. Rev. Lett. {\bf 51} (1983) 1415 and Phys. Rev.
{\bf D32} (1985) 2988;
L. Krauss, J. Moody, F. Wilczek and D. Morris, Phys. Rev. Lett.
{\bf 55} (1985) 1797.
\item{38.} S. DePanfilis et al., Phys. Rev. Lett. {\bf 59} (1987) 839
and Phys. Rev.
{\bf D40} (1989) 3153; C. Hagmann et al., Phys. Rev. {\bf D42} (1990) 1297.
\item{39.} K. van Bibber et al.,  ``Status of the Large-Scale
Dark-Matter Axion Search,'' LLNL preprint  UCRL-JC--118357,
to be published in the Proceedings of the International Conference
on {\it Critique of the Sources of Dark Matter in the
Universe}, Bel Air, CA, February 16--18, 1994.
\item{40.} D. Lynden-Bell, MNRAS {\bf 136} (1967) 101.
\item{41.} P. Sikivie and J. Ipser, Phys. Lett. {\bf B291} (1992) 288.
\item{42.} P. Sikivie, I. Tkachev and Y. Wang, to be published.
\bye